\documentclass[%
 reprint,
superscriptaddress,
 amsmath,amssymb,
 aps,prl
]{revtex4-2}

\usepackage{graphicx}% Include figure files
\usepackage{dcolumn}% Align table columns on decimal point
\usepackage{bm}% bold math
\usepackage{float}
\usepackage{lipsum}
\usepackage{hyperref}
\usepackage{color}
\usepackage[deletedmarkup=xout]{changes}

\definechangesauthor[color=red]{ZD}

\definechangesauthor[color=orange]{VM}

\definechangesauthor[color=blue]{JS}

\definechangesauthor[color=purple]{AD}

\begin{document}

%\preprint{APS/123-QED}

\title{Electron tennis driven by radio-frequency electric fields in \\low-pressure plasma sources}% Force line breaks with \\
%\thanks{A footnote to the article title}%

\author{M\'at\'e Vass}
 \affiliation{Department of Electrical Engineering and Information Science, Ruhr-University Bochum, D-44780, Bochum, Germany}
  \affiliation{Institute for Solid State Physics and Optics, Wigner Research Centre for Physics, H-1121 Budapest, Konkoly-Thege Mikl\'os str. 29-33, Hungary}
\author{Aranka Derzsi}%
 \affiliation{Institute for Solid State Physics and Optics, Wigner Research Centre for Physics, H-1121 Budapest, Konkoly-Thege Mikl\'os str. 29-33, Hungary}
 \author{Julian Schulze}
 \affiliation{Department of Electrical Engineering and Information Science, Ruhr-University Bochum, D-44780, Bochum, Germany}
\affiliation{ Key Laboratory of Materials Modification by Laser, Ion, and Electron Beams (Ministry of Education), School of Physics, Dalian University of Technology, Dalian 116024, People's Republic of China }
 \author{Zolt\'an Donk\'o}%
 \affiliation{Institute for Solid State Physics and Optics, Wigner Research Centre for Physics, H-1121 Budapest, Konkoly-Thege Mikl\'os str. 29-33, Hungary}
 \email{vass@aept.ruhr-uni-bochum.de}

\date{\today}% It is always \today, today,
             %  but any date may be explicitly specified

\begin{abstract}
We present a detailed analysis of electron trajectories within the sheath regions of capacitively coupled plasmas excited by radio-frequency voltage waveforms at low pressures. Complex features inside the sheaths are identified in several physical quantities, which are sculptured by the trajectories of the bouncing electrons under the influence of the spatio-temporally varying electric field. A method is developed to explain the generation of the various features as a function of surface processes and to identify the trajectories of electrons of different origin. 
\end{abstract}

%\keywords{Suggested keywords}%Use showkeys class option if keyword
                              %display desired
\maketitle

In low temperature plasmas, several types of neutral and charged particle populations co-exist, which under most conditions possess remarkably different distribution functions \cite{liebermanbook}. In most cases only the neutrals can be characterised by a {\it temperature}, as the distribution functions of the charged species (electrons and ions) usually deviate from the Maxwell-Boltzmann (MB) type, forbidding the use of the concept of temperature \cite{basti1}. Consequently, these feebly ionized gaseous systems exist in the state of pronounced {\it thermodynamic non-equilibrium}. The electron population at low pressures is especially weakly coupled to the neutral population and exhibits typically a large deviation from MB statistics \cite{Tsendin_2010}. Besides the non-equilibrium nature of these systems, their physics is further enriched by the {\it non-local} character of the electron transport: in typical low-pressure settings the electric field usually varies in space along the free flights of the electrons, or varies in time between two collisions. The presence of boundaries gives rises to an additional complication as it distorts the velocity distribution function \cite{kag1}. These peculiarities make low-temperature plasmas a perfect playground for kinetic theory. 

Approaches being capable of capturing  {\it kinetic effects} include the solution of the Boltzmann equation and particle based simulations \cite{verb, tejero}. The understanding of the kinetics of the electrons in these systems is of primary interest, as these particles are principally responsible for sustaining the plasma. They can be ``born'' both in gas-phase processes and at surrounding surfaces and can interact with these surfaces subsequently in different ways, such as elastic and inelastic reflection, and cause emission of additional electrons \cite{benedek2}. 
Depending on the electron kinetics
distinct operation modes (such as the $\alpha$- and $\gamma$-modes) and transitions between these exist in radio-frequency (RF) plasmas. The $\alpha$-mode is associated with electron heating at the edge of the expanding sheath, while in the $\gamma$-mode secondary electrons (``$\gamma$-electrons'') emitted from the electrodes and accelerated by the sheath electric field, cause significant ionization and increase the plasma density \cite{boeuf}. Mode transitions are intimately coupled with the electron energy distribution  function \cite{kim}. 

\begin{figure}[H]
\centering
\includegraphics[width=.48\textwidth]{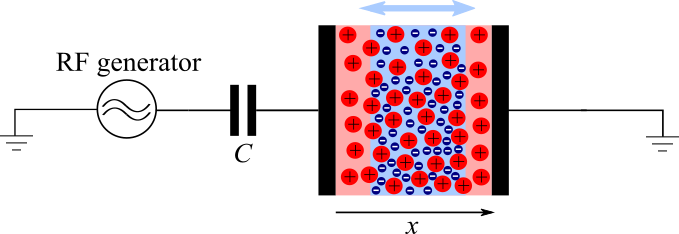}% Here is how to import EPS art
\caption{\label{fig:sketch} Scheme of the system under investigation. The plasma is established by coupling the power capacitively into the system: electrons follow the fast variation of the electric field, whereas ions react to the time averaged electric field only. The system splits into sheath regions adjacent to the electrodes and a quasineutral bulk region in the center.}
\end{figure}

Except of the secondaries, electrons are widely considered to reside outside the sheath regions, which connect the bulk plasma and the bounding surfaces, see Fig. \ref{fig:sketch}. This region has been studied intensively \cite{kag2,rpbsheath,czarnetzki}, as the proper understanding of the interaction between electrons and the sheath is of paramount importance for solving critical problems in low temperature plasma science, e.g. the lack of insights into electron power absorption. Studies of the electrons' power absorption (termed traditionally as ``electron heating'') have a long history \cite{godyak1}. The Hard Wall model, e.g., attributes energy gain to the interaction of electrons with the moving sheath boundaries \cite{liebermancikk}. This model assumes that the field is present only within the sheaths and electrons are present only outside the sheaths. The boundaries act like tennis rackets, which, in this model, have infinitely stiff strings. In reality, high-energy electrons accelerated through the sheath at one electrode can penetrate into the opposite sheath until reflected by the field, like by a tennis racket with elastic strings. The model introduced by Turner et al., called {\it pressure heating}, can also be understood based on this picture: the amount of energy gained/lost by the electrons by interacting with the sheath can be nonzero on time average due to the temporal variation of the sheath electric field during the time when the electron is present in the sheath region \cite{turner1, turner2,turner3}.

\begin{figure}[H]
\centering
\href{https://drive.google.com/file/d/18BR9LB_yAhTxgcZ7iA6AmAGTcZHjvpSJ/view?usp=sharing}
          {\includegraphics[width=.48\textwidth]{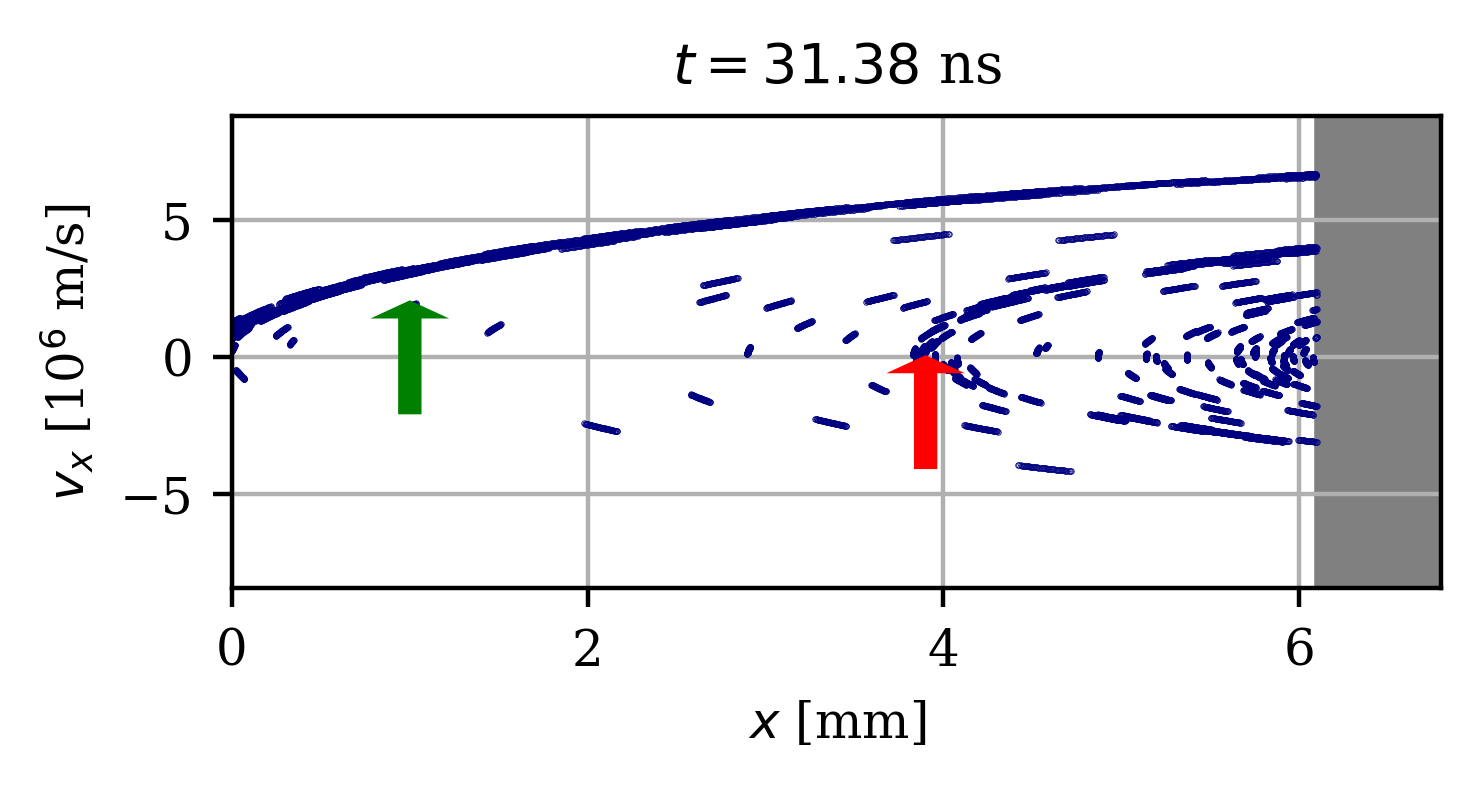}}
% Here is how to import EPS art
\caption{\label{fig:traj}  Configuration space of the electrons at $t=31.38$~ns during sheath expansion within the RF-cycle. The powered electrode is at $x$ = 0 and the  grey region marks the plasma bulk. Note (i) the parabolic line  from the origin that corresponds to $\gamma$-electrons emitted from the electrode and fly towards the center (marked by the green arrow) and (ii) the pronounced turning point of a {\it group} of electrons at $\approx$ 4 mm inside the sheath (marked by the red arrow). The ``traces'' of individual electrons correspond to a time window of $\approx$ 0.14 ns. (A movie, covering the full RF period is available as supplementary material \cite{movie}.)}
\end{figure}

Penetration of high energy electrons deep into the sheath from the plasma side is clearly revealed in Fig.~\ref{fig:traj} that shows electron trajectories in the $(x-v_x)$ configuration space in an argon plasma at $p$ = 1 Pa, $L=50$ mm electrode gap, established by an RF voltage $\Phi(t)=\Phi_0\cos(2\pi ft)$ with an amplitude of $\Phi_0=150$ V, at a frequency of $f=13.56$ MHz. These conditions are referred to as the ``base case'' in the following. The results are obtained from Particle-in-Cell / Monte Carlo collisions simulation of the discharge. The model assumes two constant surface coefficients: an elastic electron reflection probability, $R$=0.2, and an ion-induced secondary electron emission coefficient, $\gamma = $ 0.4.

At low pressures, the electrons interacting with one sheath can reach the opposite sheath without undergoing many collisions, thus leading to a scenario where electrons ``bounce'' back and forth between the two sheaths and {\it Bounce Resonance Heating} can occur \cite{brh1}. As in reality there is an electric field even outside the sheath region, predominantly the so-called {\it ambipolar} electric field \cite{julianprl,SchulzeAmb}, low energy electrons, which are usually ``trapped'' within the discharge, can effectively be ``heated'' under such resonant conditions by the ambipolar field as well \cite{park}. The spatio-temporally resolved non-local dynamics of electrons, especially that of secondary electrons, at low pressures as a consequence of their interaction with boundary surfaces and with the RF modulated sheaths is not understood. 
To complete this picture a methodology that can separate different groups of electrons and can identify their origin is needed. Due to the kinetic nature of the effects involved we do not expect analytical models to be able to achieve this goal, therefore we turn to particle based simulations to incorporate this methodology. Using this approach, in this Letter we  illustrate how complex the electrons' motion can be even in a simple plasma source that consists of two plane and parallel electrodes and is excited by an RF-voltage in an atomic gas. The results reveal the complex intra-sheath dynamics of electrons and, therefore, represent an essential step towards solving the critical problem of the lack of understanding the dynamics of energetic electrons in low temperature plasmas, which is vital for knowledge based plasma process development for various applications of broad and strong societal impact, e.g. semiconductor manufacturing, plasma medicine, and agriculture.

 \begin{figure*}%[H]
\centering
\includegraphics[width=\textwidth]{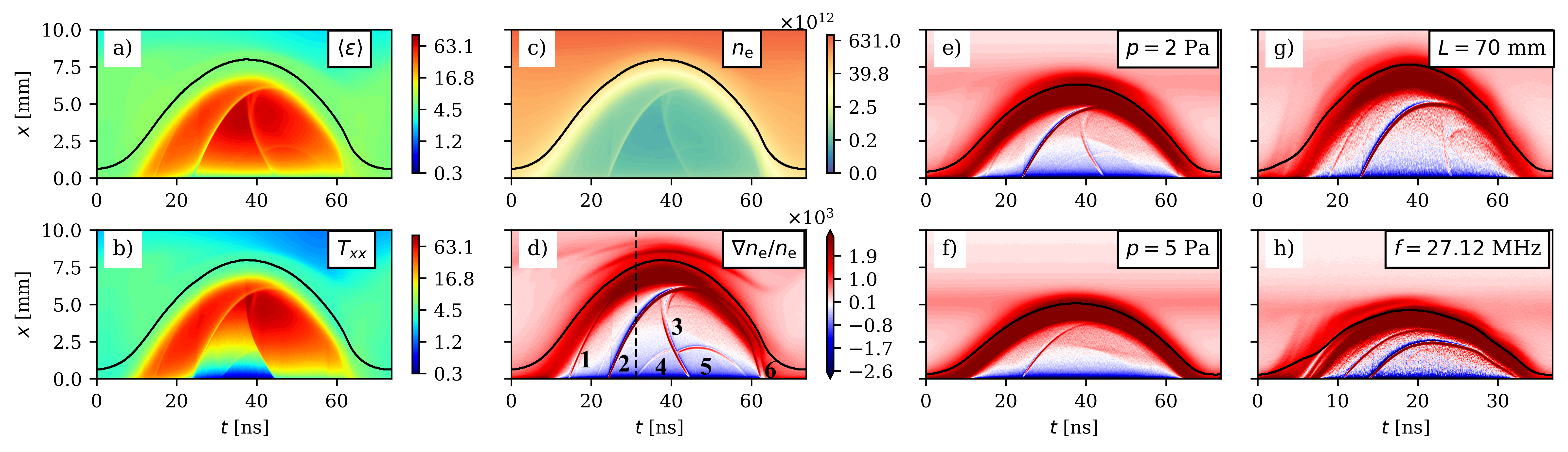}% Here is how to import EPS art
\caption{Mean energy [eV], $\langle \varepsilon\rangle$, (a), longitudinal electron temperature, $T_{xx}$ [eV], (b), density [m$^{-3}$] (c) and normalised density gradient [m$^{-1}$] (d) of the electrons for the base case ($p=1$ Pa, $L=50$ mm, $f=13.56$ MHz), and the effects of parameter variations on the normalised electron density gradient (e)-(h). The parameter that differs from the value in the base case is noted in the upper right corners of the panels. The dashed black line in panel (d) corresponds to the time instance shown in Fig.~\ref{fig:traj}. The solid black lines indicate the sheath edge \cite{rpbsheath}. The plots show only the vicinity of the powered electrode situated at $x$ = 0. The color scale of panel (d) applies for panels (e)-(h).
}
\label{fig:f23}
\end{figure*} 

%\begin{figure}[H]
%\centering
%\includegraphics[width=.48\textwidth]{Fig4.png}% Here is how to import EPS art
%\caption{\label{fig:comp} Mean energy, $\langle \varepsilon\rangle$, (a), longitudinal electron temperature, $T_{xx}$, (b) (both in units of eV), density (c) (in units of m$^{-3}$) and normalised density gradient (d) (in units of m$^{-1}$) of the electrons. The black dashed line in panel (d) corresponds to the time instance shown in figure \ref{fig:traj}. The solid black lines indicate the sheath edge \cite{rpbsheath}. The plots show only the vicinity of the powered electrode situated at $x$ = 0.}
%\end{figure}

The penetration of the fast electrons into the sheaths -- made possible by a ``soft'' boundary -- gives rise to various ``arc-like'' features that can be observed when analysing the spatio-temporal distribution of several physical quantities related to the electrons, as illustrated in Figs. \ref{fig:f23}(a)-(d) for the base case specified above: the mean energy, $\langle\varepsilon\rangle$ (a), longitudinal temperature, $T_{xx}$ (b), density, $n_{\rm e}$ (c) and normalised density gradient, $\nabla n_{\rm e}/n_{\rm e}$ (d). The mean energy and density show very similar structures within the sheath together with the longitudinal temperature, whereas in the latter there is a difference in the region between $\approx 20-45$ ns in the RF-cycle (near the electrode). The reason for this is, that in this temporal domain only a few electrons, coming from the center  of the discharge, can penetrate deep into the sheath. Thus, predominantly $\gamma$-electrons (emitted from the adjacent electrode) will contribute to the mean energy and the temperature. The definition of the temperature is $T_{xx}=m_e(\langle v_x^2\rangle-u_x^2)$, where $m_e$ is the electron mass, $v_x$ and $u_x$ are the velocity of an electron and the mean velocity of the electrons in the $x$-direction, respectively. Thus, it is proportional to the variance of the electrons' velocity, which, for the $\gamma$-electrons, is low, as they are emitted from the electrode with the same energy and need many collisions before their velocities are randomised. This, on the other hand, does not mean, that their mean energy, defined by $\langle\varepsilon\rangle=\frac{m_e}{2}\langle v_x^2+v_y^2+v_z^2 \rangle$, cannot be large.

  \begin{figure*}%[H]
\centering
\includegraphics[width=.9\textwidth]{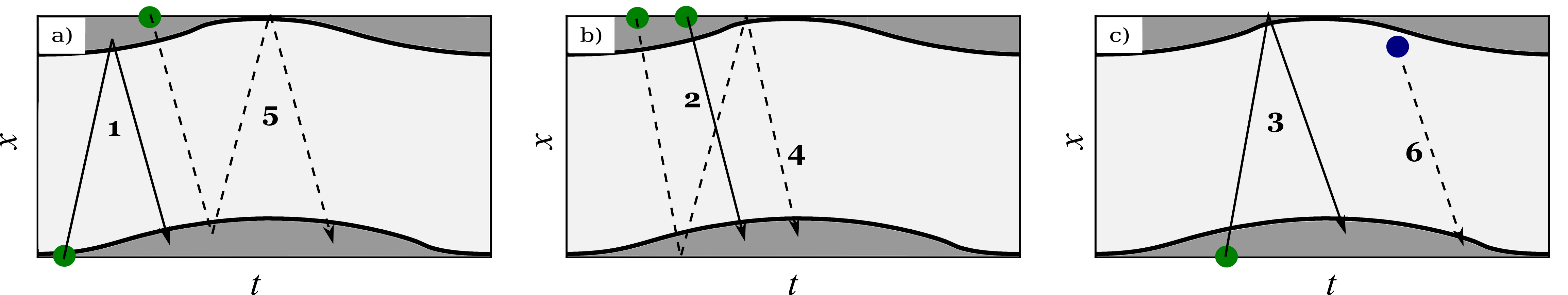}% Here is how to import EPS art
\caption{\label{fig:rxpl} Sketch of principal electron trajectories aiding the explanations of the features identified and enumerated in Fig. \ref{fig:f23}(d). Green circles denote $\gamma$-electrons, blue circles stand for bulk electrons. The powered and grounded electrodes of the discharge are situated at the bottom and at the top of the graphs, respectively. Dark gray regions: sheath domains. Light grey regions: bulk. Thick black lines: sheath edges.
}
\end{figure*}  
    
As the normalised electron density gradient shows the arc-like features most clearly, we will concentrate on this quantity from now on to identify the reason for the appearance of the different structures. The enumerated features (arcs) identified in Fig. \ref{fig:f23}(d) are the ones to be explained.
As a first step, the sensitivity of these structures on the operating conditions is addressed, as  shown in Figs. \ref{fig:f23}(d)-(f). Panel (d) corresponds to the base case. As the pressure is increased (panels (e)-(f)), the features gradually disappear: first 4, 5 and 6 already at 2 Pa, and then  1 and 3 with only feature 2 remaining at 5 Pa. As ultimately these arcs are caused by the motion of electrons, the increase in pressure means that electrons undergo more collisions and, thus, their ``collective'' motion is inhibited at higher pressures. In other words, the complex features are caused by electrons reaching the vicinity of the powered electrode, which is less likely to happen when the pressure is increased. By increasing the gap length, $L$ (g), the features get shifted in time, due to the fact, that the electrons need more time to traverse a longer gap. The frequency variation (h) results in a smaller sheath width, which is attributed to the higher plasma density. Arcs 1 and 2 can be identified in panel (h), but the others are not present. This, as we shall see,  is due to the fact that in order for a given trajectory to be formed, it matters in which phase of the sheath expansion/collapse the electron arrives at the instantaneous sheath edge.

Based on this, one can infer that the complex features in the figures above do not correspond to individual electron trajectories, but rather to the envelope of these, i.e. it is the ensemble of turning points of incoming electrons (cf. Fig. \ref{fig:traj}) which together give rise to the features. This necessitates two remarks: (i) Given this predicate, it is more understandable why we see a difference between Figs. \ref{fig:f23}(a) and \ref{fig:f23}(b): As many electrons that reach the sheath region will turn back giving rise to feature 2 in panel (d), in the region below feature 2 predominantly $\gamma$-electrons will be present, which, as elucidated above, give rise to a smaller electron temperature. 
(ii) The features identified in the normalised electron density gradient shown in Fig. \ref{fig:f23}(d) all have a negative (blue) and a positive (red) side. This can be explained by the fact, that at the turning points of the electrons, a local bunching effect happens, increasing the electron density along the features (as seen in Fig. \ref{fig:f23}(d)), and consequently, causing a positive edge at the side of the powered electrode, and a negative one in the other direction.

\begin{figure*}%[H]
\centering
\includegraphics[width=\textwidth]{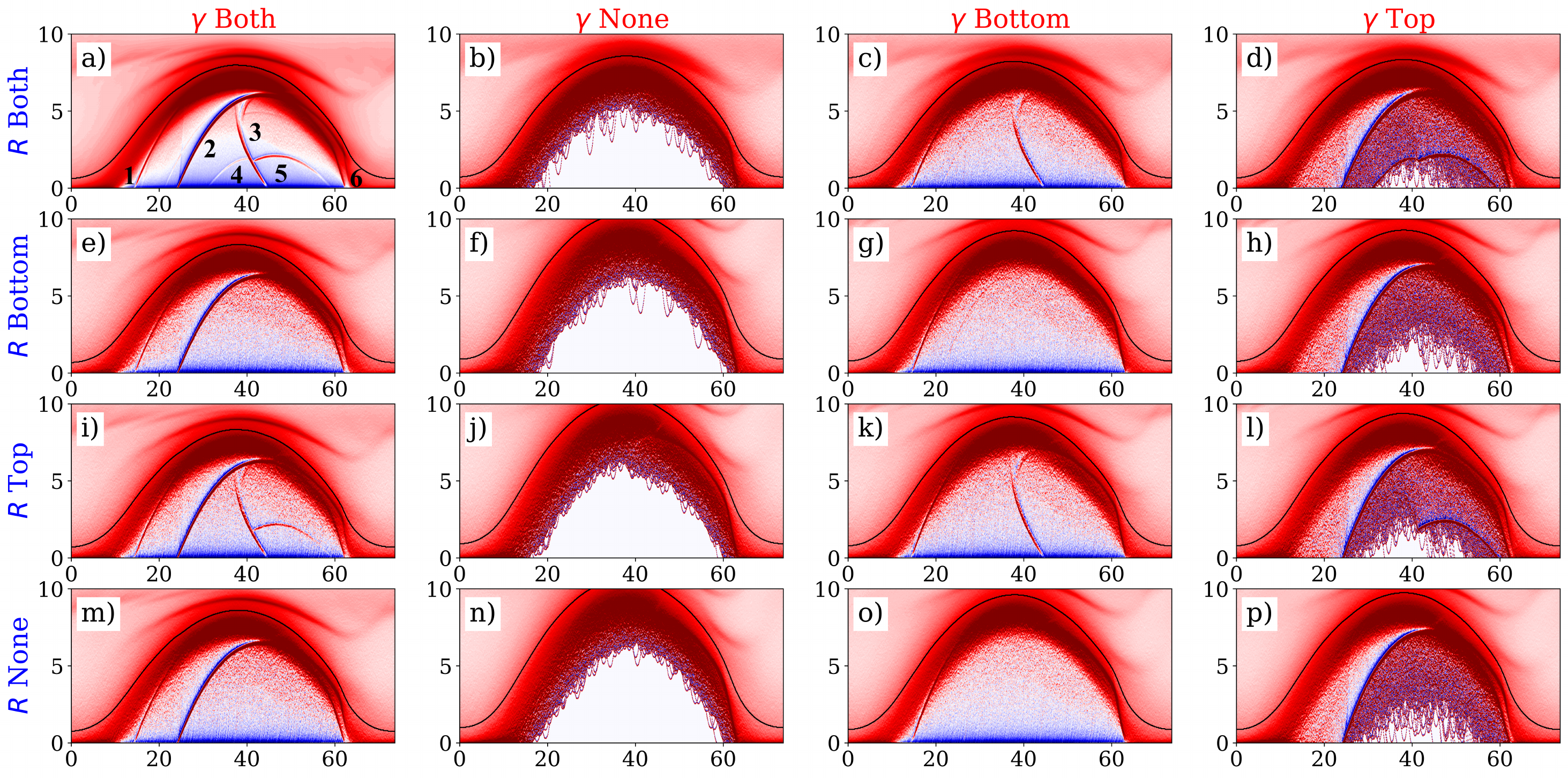}% Here is how to import EPS art
\caption{Features under different electrode surface conditions. The presence/absence of the electron reflection process (rows) and the secondary electron emission process (columns) are indicated by the keywords 'Both', 'None' 'Top' and 'Bottom', which refer to the electrodes. The bottom electrode is situated at $x=0$. For clarity, axis labels and colorbars are omitted, they agree with that of Fig. \ref{fig:f23}(d). }
\label{fig:All}
\end{figure*}

As $\gamma$-electrons reach much higher energies than the electrons born outside the sheath, they are the primary suspects for the generation of the complex features. Following their trajectories aids the explanations of these features. A sketch of the principal electron trajectories is shown in Fig. \ref{fig:rxpl}: green circles denote $\gamma$-electrons, blue circles bulk (i.e. ``born in the bulk'') electrons. Whenever an electron reaches the electrode surface, it undergoes a reflection from the electrode surface with a probability $R$. The ends of the trajectories (the arrowheads) correspond to the approximate place/time where the electrons turn back inside the sheath at the powered (bottom) electrode. 

Our method is to run simulations with every possible combination of the surface processes: we can have secondary electron emission on either electrodes, both or none (controlled by the $\gamma$-coefficient, having a value of 0 or 0.4), and similarly with the elastic electron reflection (controlled by the $R$ coefficient, having a value of 0 or 0.2). For brevity and notational simplicity, `Bottom' and `Top' electrode will be used instead of powered/grounded. The results obtained this way for the base case are shown in Fig. \ref{fig:All}.

\begin{table}%[H]
\caption{\label{tab:table1}
Appearances of the various features. }
\begin{ruledtabular}
\begin{tabular}{ccccc}
 &{\color{red} $\gamma$ Both}&{\color{red} $\gamma$ None} &{\color{red} $\gamma$ Bottom}& {\color{red} $\gamma$ Top}\\
\hline
{\color{blue} $R$ Both}& 1-6 & 6 & 1, 3, 6 &2, 4, 5, 6\\
{\color{blue} $R$ Bottom}& 1, 2, 6 & 6& 1, 6 &2, 6 \\
{\color{blue} $R$ Top}& 1, 2, 3, 5, 6 & 6 & 1, 3, 6 &2, 5, 6 \\
{\color{blue} $R$ None}& 1, 2, 6 & 6 & 1, 6 &2, 6 \\
\end{tabular}
\end{ruledtabular}
\end{table}

Table \ref{tab:table1} summarizes the appearances of the various features, by listing the numbers assigned to the features in Figs. \ref{fig:f23}(d) and \ref{fig:All}(a). Based on this table, it can readily be identified what surface process on either electrode is needed to generate the given pattern. In principle, this approach can be generalised to more than two surface processes, including e.g., secondary electron emission due to electron bombardment of the electrodes  \cite{benedek,benedek2}.

Invoking Fig. \ref{fig:rxpl}, the features are identified as follows:
\begin{itemize}
    \item Feature 1 (Fig.~\ref{fig:rxpl}(a)) is present only in the $\gamma$ Bottom column, i.e. it is caused by secondary electrons born at the bottom electrode. As the feature is present irrespective of the presence of electron reflection, it means, that a $\gamma$-electron born at the bottom electrode traverses the whole discharge, is reflected by the opposing (top) sheath and flies back to the bottom electrode again. This can happen, as taking a $\gamma$-electron having an average energy of $\approx75$ eV, it can traverse the gap of 50 mm in $\approx10$ ns. 
    \item Feature 2 (Fig.~\ref{fig:rxpl}(b)) - as it is present in the $\gamma$ Top column, it is caused by $\gamma$-electrons born at the top electrode, accelerated by the sheath electric field and reaching the bottom sheath where they are turned back.
    \item Feature 3 (Fig.~\ref{fig:rxpl}(c)) - as it is present in the $\gamma$ Bottom and $R$ Top panels, thus, the feature is proven to be generated by those $\gamma$-electrons, which are born at the bottom electrode, traverse the discharge, reach the opposite electrode where they get reflected, and come back to the bottom sheath again.
    \item Feature 4 (Fig.~\ref{fig:rxpl}(b)) is not among the panels discussed before, which means, that it has to be a multiple reflection-caused feature. It can only be found in the $\gamma$ Top column with both electrodes having nonzero reflection coefficients ($R$ Both row). This means, that this feature is caused by electrons born at the top electrode, reflected twice by either electrodes before turning back in the bottom sheath region.
    \item Feature 5 (Fig.~\ref{fig:rxpl}(a)) is again a multiple reflection-caused feature, because it is found in the $\gamma$ Top column and the $R$ Top row, which means the electron is born at the top electrode, but also reflected by it. The only way it can happen is, if it is first reflected by the bottom sheath and then the top electrode. 
    \item Feature 6 (Fig.~\ref{fig:rxpl}(c)) is the only structure which appears in the $\gamma$ None column, and is not sensitive to reflections. Thus, it is caused by electrons generated within the plasma bulk that reach the bottom sheath.
\end{itemize}

In summary, the complex dynamics of fast electrons was analysed via simulations in a low pressure capacitively coupled plasma. Prominent features in a number of various physical parameters were found inside the sheath regions. We have successfully associated these features with the distinct electron creation mechanisms and the motion of the electrons in the spatio-temporally varying electric field. It was inferred that the features do not correspond to trajectories, but are envelopes of the turning points of these electrons within the sheath region. The determination of the underlying mechanisms leading to the specific features was done by a procedure where the electrode surface processes were turned on/off (by assuming nonzero/zero values for the surface coefficients) and all possible combinations have been scanned. This new methodology enables the understanding of the complex and non-local spatio-temporal dynamics of energetic electrons in technological plasmas. The results obtained represent an important step forward to solving this outstanding problem, which is of broad fundamental and applied relevance due to the manifold of applications of high societal relevance of these plasma sources. 

The authors acknowledge funding from the German Research Foundation in the frame of the project, ``Electron heating in capacitive RF plasmas based on moments of the Boltzmann equation: from fundamental understanding to knowledge based process control'' (No. 428942393), via SFB TR 87, project C1, and from the Hungarian National Office for Research, Innovation, and Technology (NKFIH), via grants K119357 and FK128924.

% The \nocite command causes all entries in a bibliography to be printed out
% whether or not they are actually referenced in the text. This is appropriate
% for the sample file to show the different styles of references, but authors
% most likely will not want to use it.
%\nocite{*}

\providecommand{\noopsort}[1]{}\providecommand{\singleletter}[1]{#1}%

\end{document}